\documentclass[aps,pra,showpacs,twocolumn]{revtex4-1}
\usepackage{amssymb}
\usepackage{amsmath}
\usepackage{graphicx}
\usepackage{epsfig}

\begin{document}

\title{Generation of Bell, W and GHZ states via exceptional points in
non-Hermitian quantum spin systems}
\author{C. Li and Z. Song}
\email{songtc@nankai.edu.cn}
\affiliation{School of Physics, Nankai University, Tianjin 300071, China}

\begin{abstract}
We study quantum phase transitions in non-Hermitian XY and transverse-field
Ising spin chains, in which the non-Hermiticity arises from the imaginary
magnetic field. Analytical and numerical results show that at exceptional
points, coalescing eigenstates in these models close to W, distant Bell and
GHZ states, which can be steady states in dynamical preparation scheme
proposed by T. D. Lee et. al. (Phys. Rev. Lett. \textbf{113}, 250401
(2014)). Selecting proper initial states, numerical simulations demonstrate
the time evolution process to the target states with high fidelity.
\end{abstract}

\pacs{11.30.Er, 03.67.Bg, 75.10.Jm}
\maketitle


\section{Introduction}

Quantum phase transition can occur in a finite non-Hermitian system,
associating parity-time ($\mathcal{PT}$) reversal or other type of symmetry
breaking. At the transition point as referred to exceptional point (EP), a
pair of eigen states coalesces into a single state. Many finite-sized
discrete systems have been investigated, including tight-binding models,
quantum spin chains, and complex crystal.

These features are different from that of quantum phase transition in
infinite Hermitian system. Recently, critical behavior of non-Hermitian
system has been employed to generate entangled states in a dynamical process
and the corresponding experimental protocol is also proposed \cite%
{Tony1,Tony2}. According to the non-Hermitian quantum theory \cite%
{Bender,Ann,JMP1,JPA1,PRL1,JMP2,JPA3,JPA5}, a pseudo-Hermitian system has
real eigenvalues or conjugate pair complex eigenvalues. Considering the
simplest case, there is only a single pair of eigenstates breaking the
symmetry of the Hamiltonian, with conjugate complex eigenvalues. A seed
state is an initial state consisting of various eigenstates with eigenvalues
with zero, positive and negative imaginary parts, respectively. As time
evolution, the amplitude of the state with positive imaginary part in its
eigenvalues will increase exponentially and suppress that of other
components. The target is the final steady state and expected to have
peculiar features for quantum computation processing and other applications.
It is important to construct a simple Hamiltonian which is suitable for
experimental implementation: to prepare desirable quantum states with high
fidelity.

In quantum information science, it is a crucial problem to develop
techniques for generating entanglement among stationary qubits, which plays
a central role in applications \cite{Ekert, Deutsch, Bennett}. Bell states
are specific maximally entangled quantum states of two qubits. For
many-qubit system, there are two typical multipartite entangled states,
Greenberger-Horne-Zeilinger (GHZ) and W states, which are usually referred
to as maximal entanglement. Multipartite entanglement has been recognized as
a powerful resource in quantum information processing and communication.
Numerous protocols for the preparation of such states have been proposed
\cite{Cirac, Gerry, Hagley, Cabrillo, Bose, Lange, Rauschenbeutel, Zheng,
Feng, Simon, Zou, Duan, SongJ, Su, SongHS, LiY PRA, Jin2}.

In this paper, we consider whether it is possible to use non-Hermitian
systems to generate a W, distant Bell and GHZ states via the dynamical
process near EPs. We introduce a non-Hermitian $XY$ and a transverse-field
Ising spin chains to demonstrate the schemes. Numerical simulations show
that the target states can be obtained with high fidelity by the time
evolutions of selecting proper initial states.

The remainder of this paper is organized as follows. In Sec. \ref{$XY$ Spin
chain}, we present a non-Hermitian $XY$ spin model and solutions. Secs. \ref%
{W state}, \ref{Bell state} and \ref{GHZ state} are devoted to the schemes
of preparing W, Bell and GHZ states, respectively. Finally, we present a
summary and discussion in Sec. \ref{Summary}.

\begin{figure*}[tbp]
\includegraphics[ bb=23 261 436 566, width=0.45\textwidth, clip]{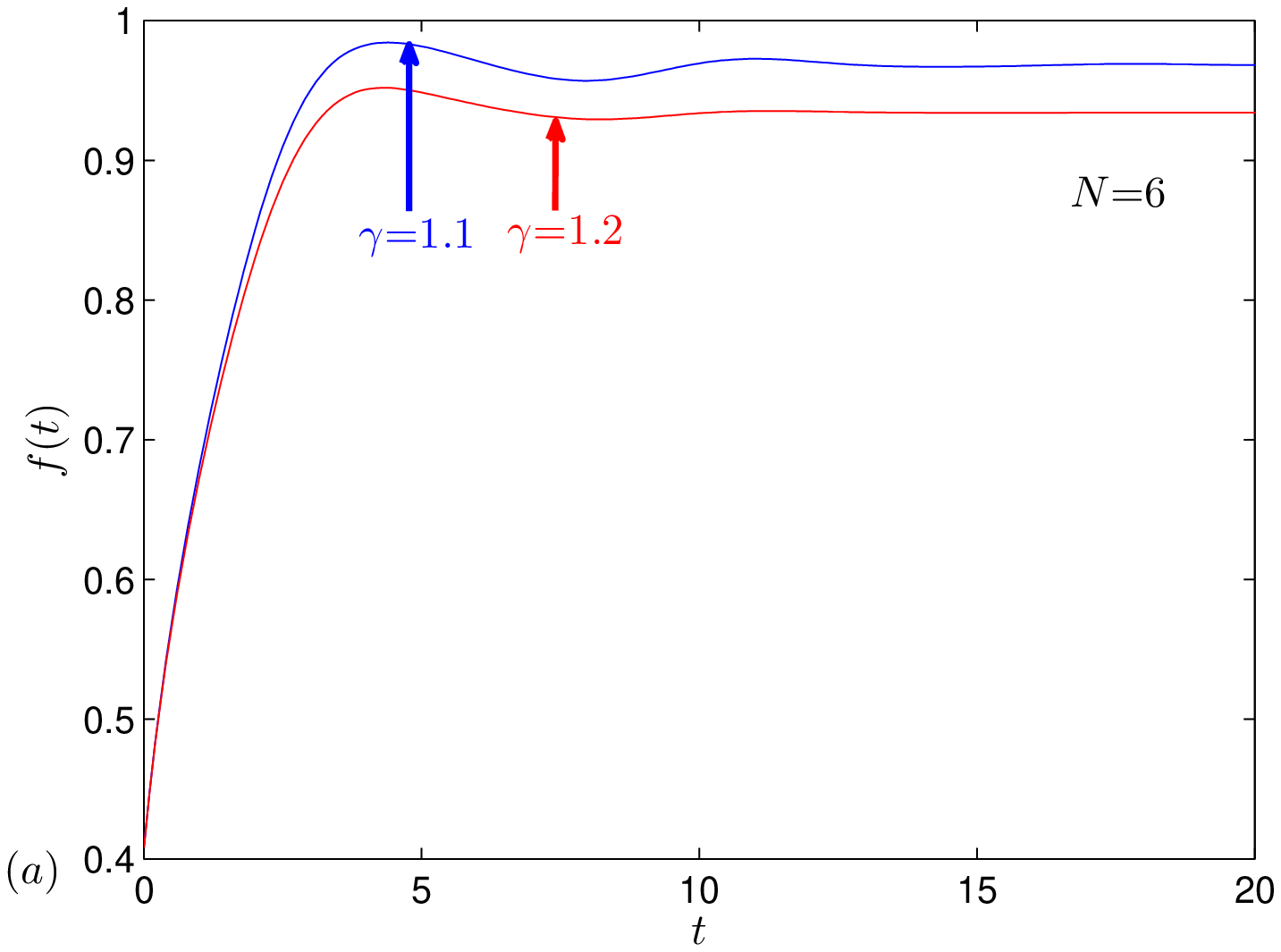} %
\includegraphics[ bb=23 261 436 566, width=0.45\textwidth, clip]{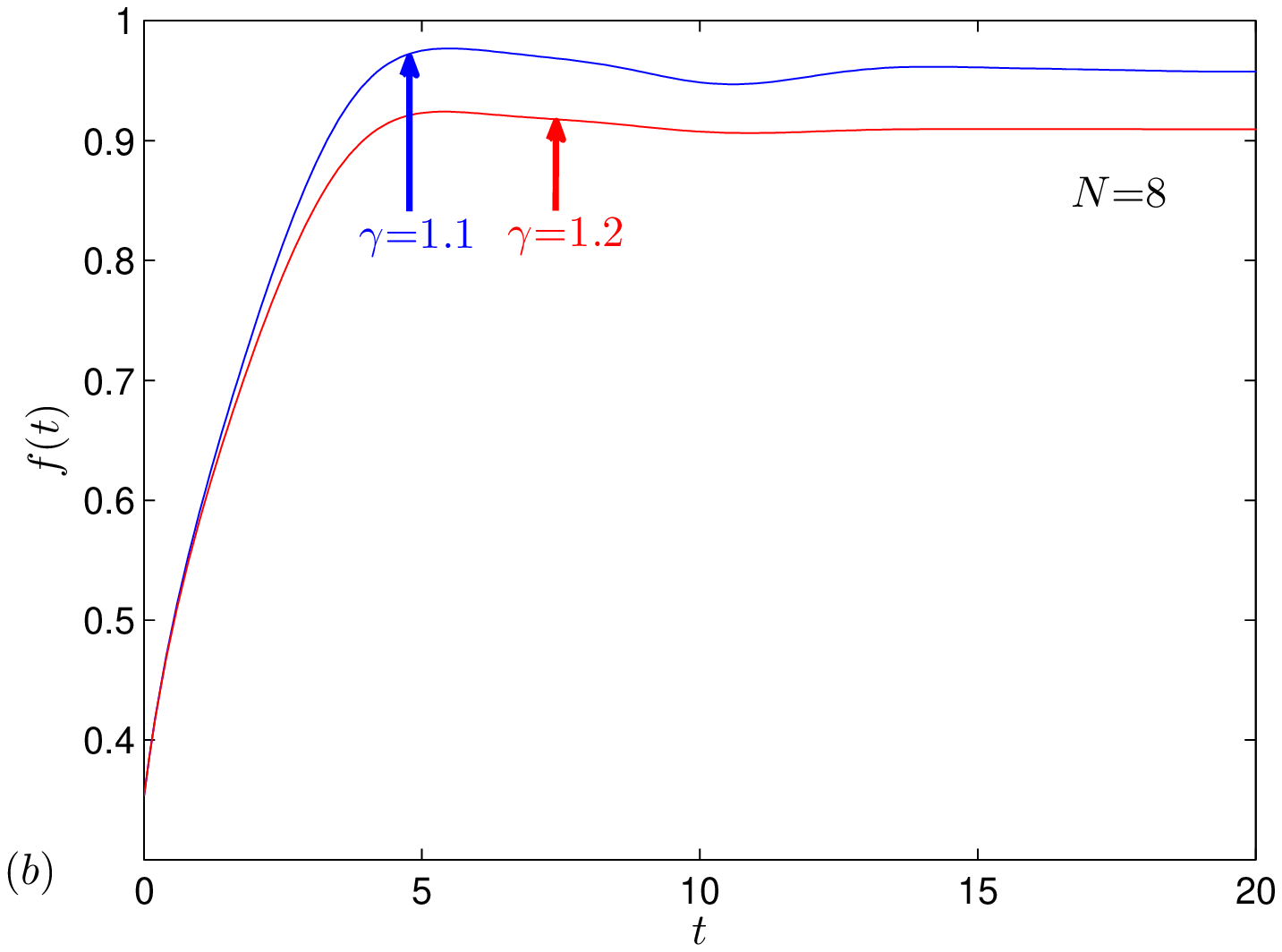}
\caption{(Color online) Plots of the fidelity $f\left( t\right) $ for
preparing W state,\ as a function of time for the systems with $N=6$ (a) and
$8 $ (b). The times are dimensionless and in units of $1$.\ We see that the
fidelities converge to constants\ in exponential manner and the converging
fidelities become higher as $\protect\gamma $ closes to $1$, while the
converging times get longer. The obtained results are not sensitive to the
size $N$, which is quite different from the situations for the productions
of Bell and GHZ states. }
\label{fig1}
\end{figure*}

\section{$XY$ Spin chain}

\label{$XY$ Spin chain}We consider a non-Hermitian $XY$ spin model
\begin{eqnarray}
H_{\text{chain}} &=&\frac{1}{2}\sum_{l=1}^{N-1}\left( \sigma _{l}^{x}\sigma
_{l+1}^{x}+\sigma _{l}^{y}\sigma _{l+1}^{y}\right) +\text{\textrm{H.c.}}
\notag \\
&&\left( V+i\gamma \right) \sigma _{1}^{z}+\left( V-i\gamma \right) \sigma
_{N}^{z},  \label{H_chain}
\end{eqnarray}%
on an $N$-site chain, where $\sigma _{l}^{\alpha }$ ($\alpha =x,y,z$) is
Pauli matrix. In the case of $\gamma =0$, it is reduced to a Hermitian model
with $\mathcal{P}$ symmetry. Here the parity operator $\mathcal{P}$ is given
by $\mathcal{P}\sigma _{l}^{\alpha }\mathcal{P}^{-1}=\sigma _{\bar{l}%
}^{\alpha }$ with $\bar{l}=(N+1-l)$. In the case of nonzero $\gamma $, the $%
\mathcal{P}$\ symmetry is broken, but $\mathcal{PT}$ is still symmetric,
where $\mathcal{T}$\ is a time reversal operator $\mathcal{T}i\mathcal{T}%
^{-1}=-i$.

We note that%
\begin{equation}
\left[ J_{z},H_{\text{chain}}\right] =0,
\end{equation}%
where $J_{\alpha }=\sum _{l=1}^{N}\sigma _{l}^{\alpha }$ is a total spin
operator. This means that $H_{\text{chain}}$ can be diagonalized in each
invariant subspace.In this paper, we only concern the issue in the subspace
with $J_{z}=N-1$ and $N=$even. In this invariant subspace, the wave function
has the form

\begin{equation}
\left\vert \phi \right\rangle =\sum_{l=1}^{N}f_{l}\sigma _{l}^{+}\left\vert
\Downarrow \right\rangle ,
\end{equation}%
where $\left\vert \Downarrow \right\rangle $ is a saturated ferromagnetic
state $\left\vert \Downarrow \right\rangle =\prod_{l=1}^{N}\left\vert
\downarrow \right\rangle $. Then we get an equivalent Hamiltonian

\begin{eqnarray}
H_{\text{eq}} &=&\sum_{l=1}^{N-1}\left\vert l\right\rangle \left\langle
l+1\right\vert +\text{H.c.}  \notag \\
&&+\left( V+i\gamma \right) \left\vert 1\right\rangle \left\langle
1\right\vert +\left( V-i\gamma \right) \left\vert N\right\rangle
\left\langle N\right\vert ,  \label{H_eq}
\end{eqnarray}%
where the position state at $l$th site is $\left\vert l\right\rangle \equiv
\sigma _{l}^{+}\left\vert \Downarrow \right\rangle $. The eigen problem of
the equivalent Hamiltonian is given in Appendix. In the following, we will
discuss the schemes for the preparation of W and Bell states based on the
Hamiltonian $H_{\text{eq}}$.

\section{W state}

\label{W state}In the situation $V=0$, the Hamiltonian $H_{\text{eq}}$\ is
reduced to%
\begin{equation}
H_{\text{W}}=\sum_{l=1}^{N-1}\left\vert l\right\rangle \left\langle
l+1\right\vert +\text{H.c.}+i\gamma \left\vert 1\right\rangle \left\langle
1\right\vert -i\gamma \left\vert N\right\rangle \left\langle N\right\vert .
\end{equation}%
The exact solution in the Appendix suggests us to consider the state%
\begin{equation}
\left\vert \text{W}\right\rangle =\frac{1}{\sqrt{N}}\sum_{l=1}^{N}\left(
-i\right) ^{l}\sigma _{l}^{+}\left\vert \Downarrow \right\rangle =\frac{1}{%
\sqrt{N}}\sum_{l=1}^{N}\left( -i\right) ^{l}\left\vert l\right\rangle ,
\end{equation}%
which represents a single-magnon spin wave with wave vector $\pi /2$.\ It is
a W state under a local transformation $\left( -i\right) ^{l}\sigma
_{l}^{+}\rightarrow \sigma _{l}^{+}$, which does not reduce its properties
in quantum information processing. A straightforward derivation shows that
the state $\left\vert \text{W}\right\rangle $\ is an eigenstate of the
Hamiltonian $H_{\text{W}}$\ at $\gamma =\gamma _{c}=1$, i.e.,
\begin{equation}
H_{\text{W}}\left( \gamma _{c}\right) \left\vert \text{W}\right\rangle
=0\left\vert \text{W}\right\rangle .
\end{equation}%
Then, we will show that $\left\vert \text{W}\right\rangle $ is a special
eigenstate of $H_{\text{W}}\left( \gamma _{c}\right) $. For the
corresponding conjugate Hamiltonian $H_{\text{W}}^{\dag }\left( \gamma
_{c}\right) $, we have%
\begin{equation}
H_{\text{W}}^{\dag }\left( \gamma _{c}\right) \left\vert \mathcal{W}%
\right\rangle =0\left\vert \mathcal{W}\right\rangle ,
\end{equation}%
where%
\begin{equation}
\left\vert \mathcal{W}\right\rangle =\frac{1}{\sqrt{N}}\sum_{l=1}^{N}i^{l}%
\sigma _{l}^{+}\left\vert \Downarrow \right\rangle
\end{equation}%
It is easy to find that%
\begin{equation}
\langle \mathcal{W}\left\vert \text{W}\right\rangle =0,
\end{equation}%
which indicates that $H_{\text{W}}$ has an EP at $\gamma _{c}$ and the W
state is the coalescent state at the transition point. In the Appendix, this
result is confirmed by an exact Bethe Ansatz analysis.

We investigate the scheme of selecting the $\left\vert \text{W}\right\rangle
$\ state by a dynamic process. From the Appendix or the previous work \cite%
{Jin1}, we find that the complex conjugate pair of energies is $\pm i2J\sinh
\kappa $ for small $\kappa $, where real number $\kappa $ obeys the equation%
\begin{equation}
\gamma ^{2}\sinh [\left( N-1\right) \kappa ]=\sinh [\left( N+1\right) \kappa
].
\end{equation}%
We note that the value of $\gamma $\ determines the gap between the complex
conjugate pair of energies, or the converging time. The initial state is
taken as $\left\vert \psi \left( 0\right) \right\rangle =\left\vert
1\right\rangle $, the evolved state $\left\vert \psi \left( t\right)
\right\rangle $\ is expected to close the target state for sufficient long
time. We employ the fidelity%
\begin{equation}
f\left( t\right) =\left\vert \left\langle \text{W}\right\vert \widetilde{%
\psi }\left( t\right) \rangle \right\vert ,
\end{equation}%
to characterize the efficiency of the scheme. Here $\left\vert \widetilde{%
\psi }\left( t\right) \right\rangle $\ is the Dirac normalized state of $%
\left\vert \psi \left( t\right) \right\rangle $ to reduce the increasing
norm of $\left\vert \psi \left( t\right) \right\rangle $. In the limit case
of $\gamma \rightarrow 1$, we will have $f\left( t\right) \rightarrow 1$ as $%
t\rightarrow \infty $. For finite $\gamma $, the time evolution of the state
is computed by numerical diagonalization in the broken symmetric region. In
order to quantitatively evaluate the fidelity and demonstrate the proposed
scheme, we simulate the dynamic processes of the W state preparation. To
illustrate the process, we plot the fidelities as functions of time for
systems with $N=6$ and $8$ in Fig. \ref{fig1}. It shows that the fidelities
converges to a steady value exponentially fast. Smaller $\gamma $
(approaches to $1$) can enhance the fidelity, while the converging time
becomes longer. Moreover, we find that the converging times for two cases
are not so sensitive to the size $N$, which is quite different from that in
following two schemes for preparing distant Bell and GHZ states. This is
because of the fact that the phase boundary is always at $\gamma =1$\ for
any even $N$. Then such a scheme is more efficient for a W-state production.

\begin{figure*}[tbp]
\includegraphics[ bb=11 254 459 582, width=0.32\textwidth, clip]{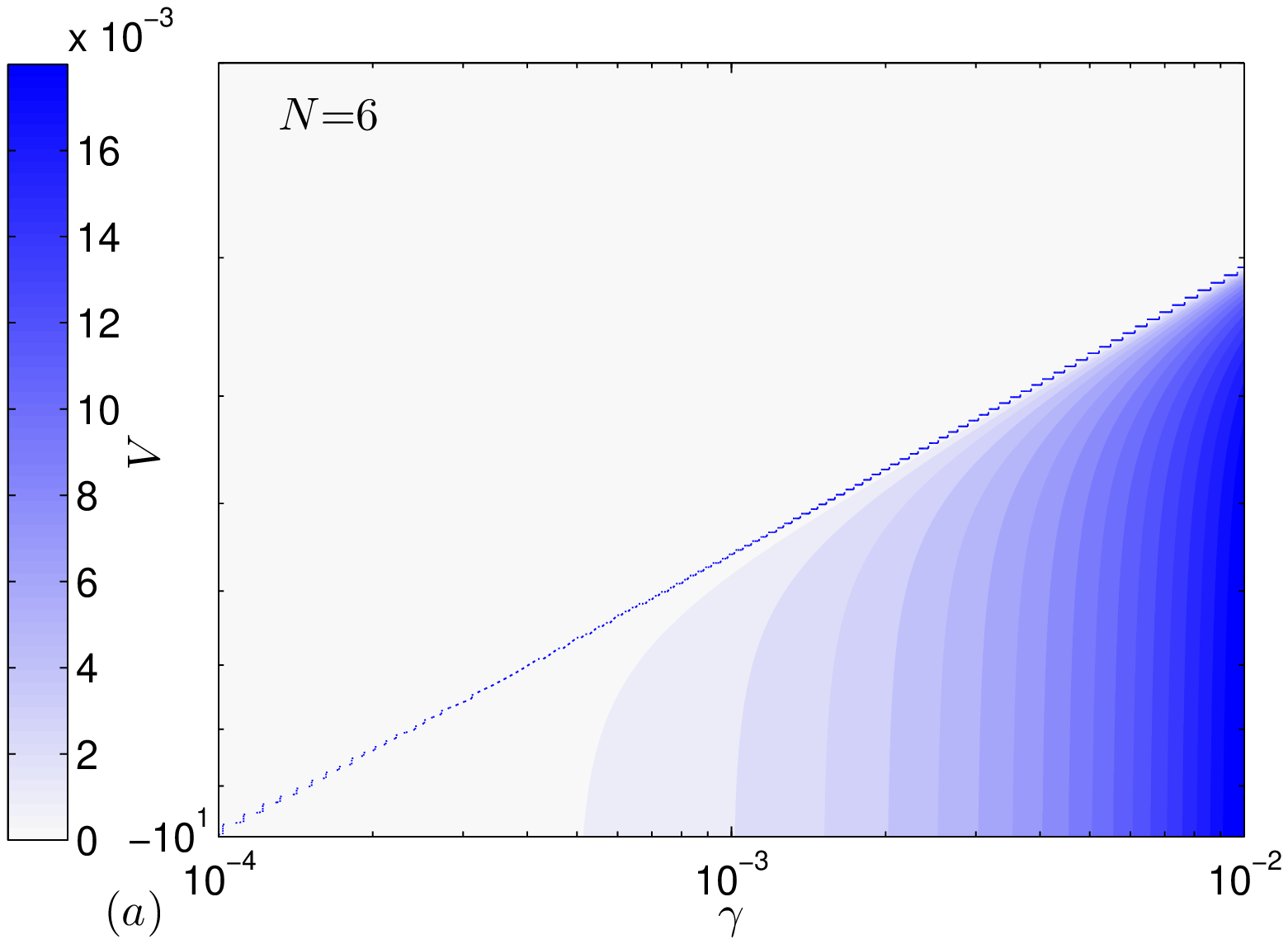} %
\includegraphics[ bb=11 252 458 582, width=0.32\textwidth, clip]{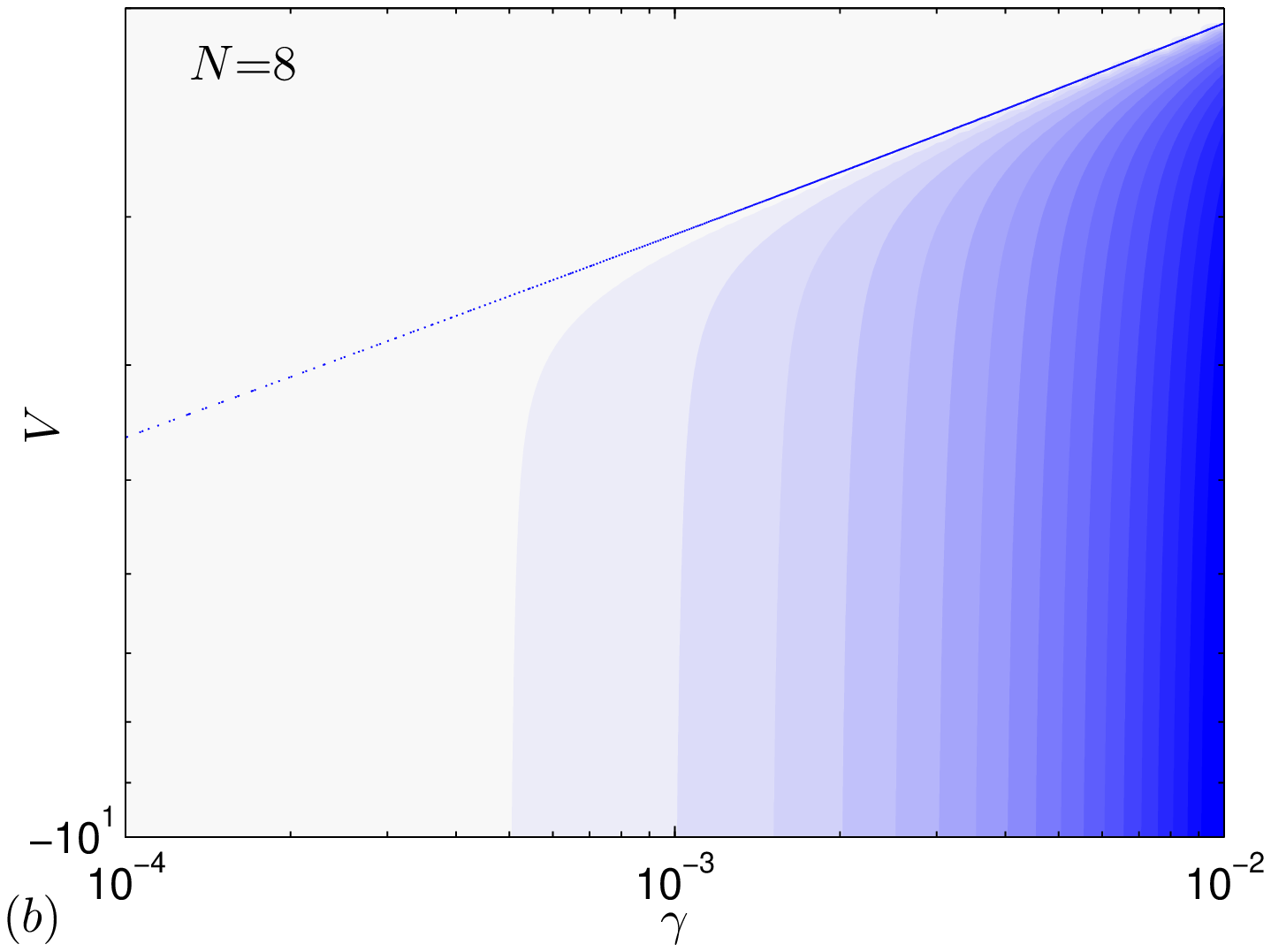} %
\includegraphics[ bb=11 252 458 582, width=0.32\textwidth, clip]{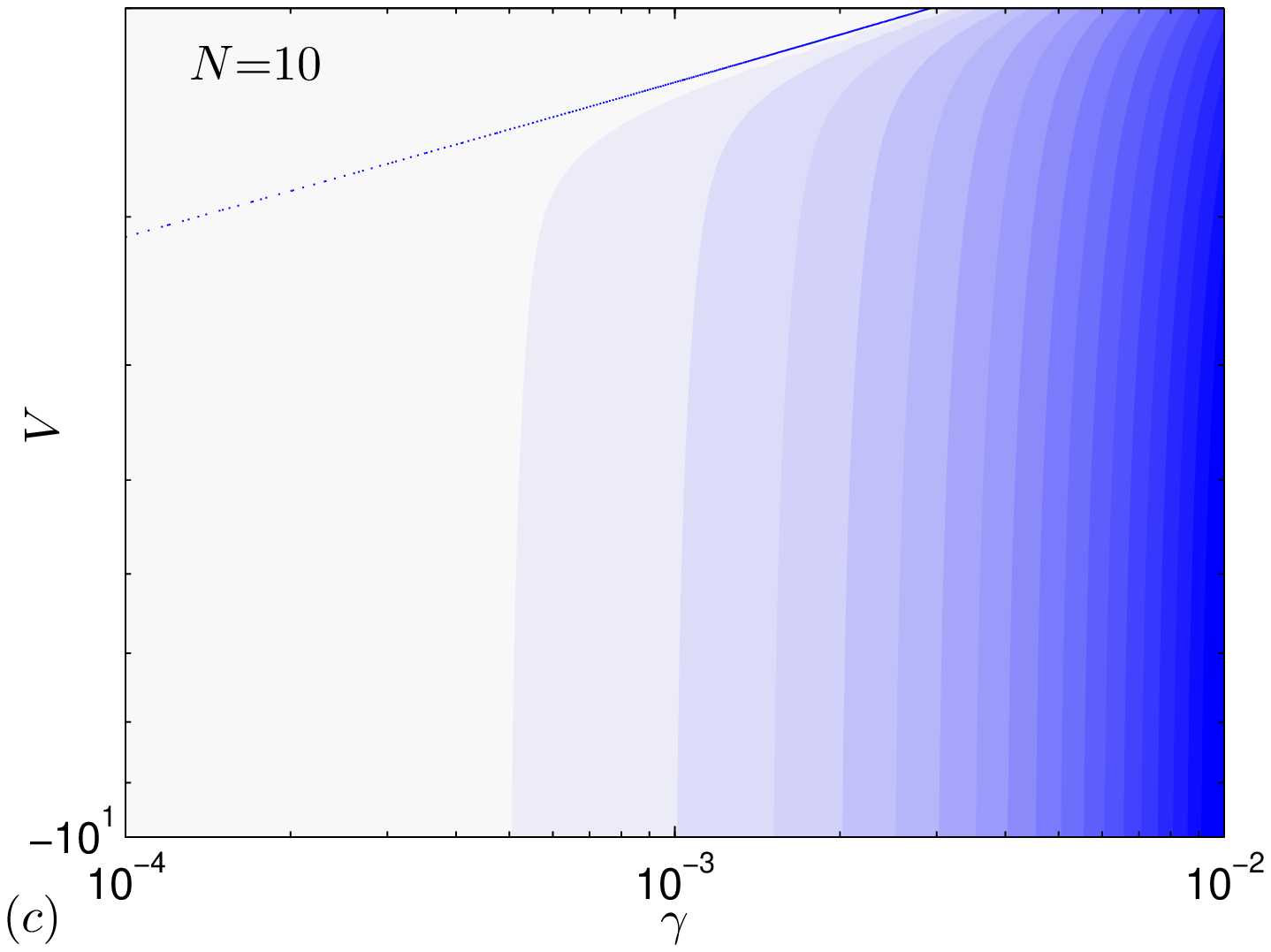}
\caption{(Color online) Phase diagram of non-Hermitian Hamiltonian in Eq. (%
\protect\ref{H_eq}). The color contour map represents the magnitude of the
imaginary part of conjugate-pair energy levels for $N=6$, $8$ and $10$,
obtained by exact diagonalization. The white area indicates the region where
the spectrum is entirely real. The dot line is the plot of function in Eq. (%
\protect\ref{EPts}), indicating the exact phase boundary. We see that all
three boundaries are in line shape with different slopes in the logarithm
scales. This property can be explained by perturbation approximation.}
\label{fig2}
\end{figure*}

\begin{figure*}[tbp]
\includegraphics[ bb=23 254 436 567, width=0.45\textwidth, clip]{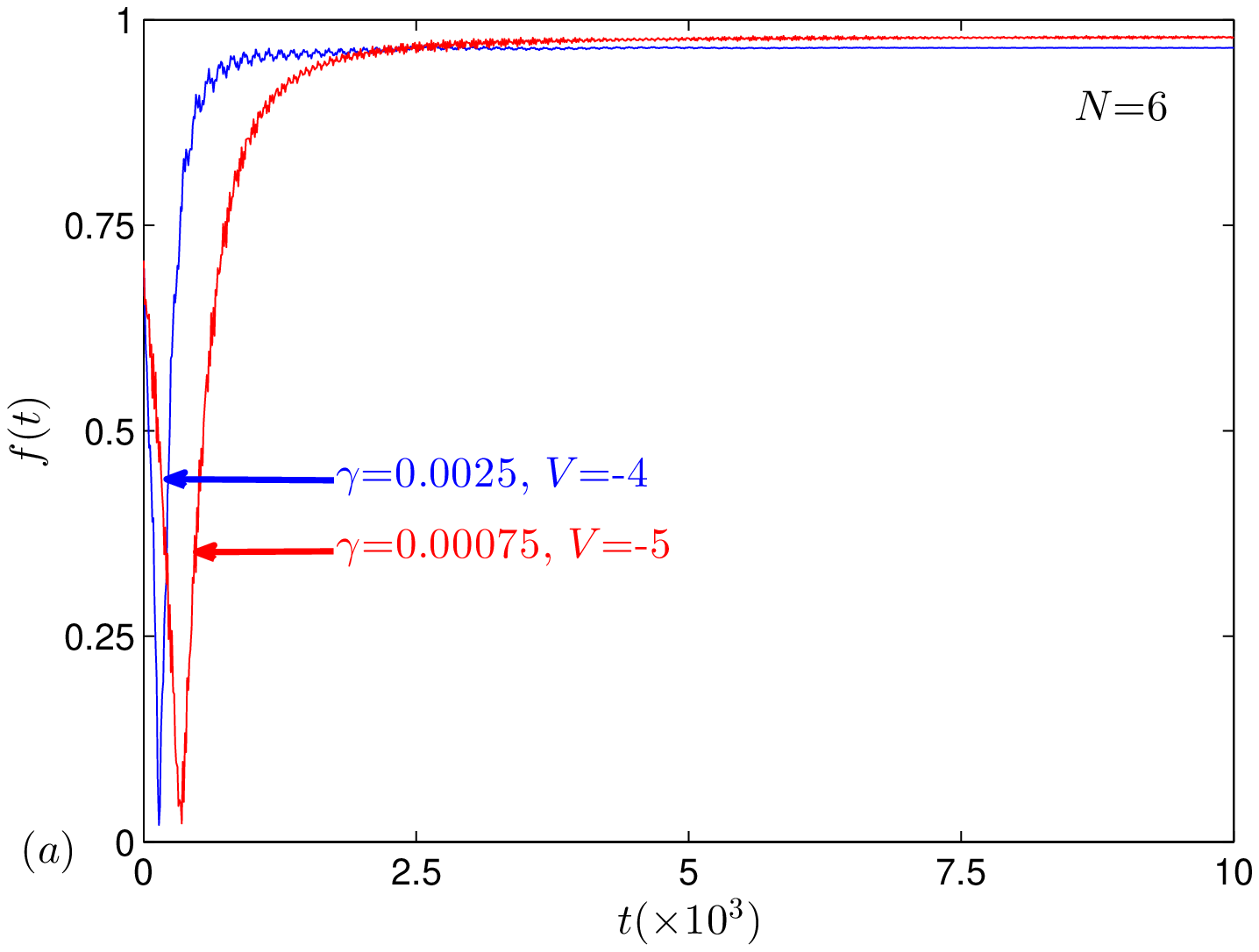} %
\includegraphics[ bb=23 254 436 567, width=0.45\textwidth, clip]{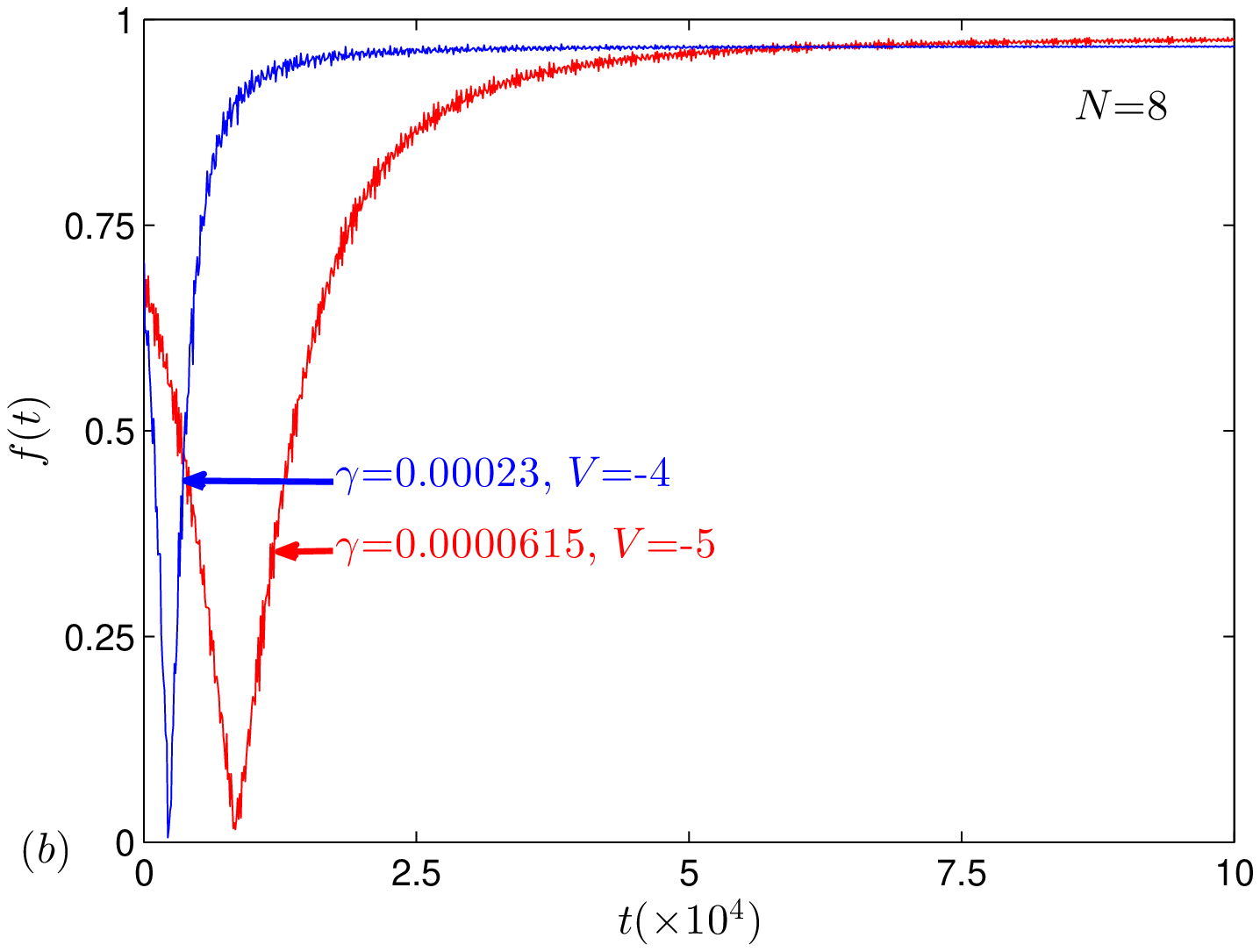}
\caption{(Color online) Plots of the fidelity $f\left( t\right) $ for
preparing Bell state,\ as a function of time for the systems with $N=6$ (a)
and $8$ (b). The times are dimensionless and in units of $10^{3}$ and $%
10^{4} $,\ respectively.\ We see that the converging fidelities approach to $%
1$, getting higher as increasing of $V$ and longer time. The time scale of
(b) is over ten times longer than that of (a), which indicates the
difficulty of preparing long-distance Bell state.}
\label{fig3}
\end{figure*}

\section{Bell state}

\label{Bell state}In the situation $\left\vert V\right\vert >2$, the exact
solution in the Appendix shows that two bound states are formed, in which
the probability mainly distributes around two ending sites. The phase
diagram has been obtained as Eq. (\ref{EPts}) in the Appendix, which is the
base of the scheme for preparing Bell state. According to the Bethe Ansatz
result, there is a conjugate complex pair of energy levels in the broken $%
\mathcal{PT}$ symmetric region. The magnitude of the imaginary part of the
eigen energy $\left\vert \text{Im}\varepsilon \right\vert $ is also an
indicator of the phase boundary and determines the converging speed of the
scheme. For illustrating this point, we plot $\left\vert \text{Im}%
\varepsilon \right\vert $ as a function of $V$ and $\gamma $\ for the
systems with $N=6$, $8$ and $10$ in Fig. \ref{fig2}. The corresponding exact
boundary from Eq. (\ref{EPts}) is plotted as well. We find that they accord
with each other and the boundary appears as a linear line with $N$-dependent
slope in the logarithm scales. We will see that the profile of the phase
diagram directly determines the efficiency of the scheme in the following
investigation.

In order to understand a clear physical picture of the exact solution, we
use the perturbation method to simplify the Hamiltonian $H_{\text{eq}}$\ in
large $V$\ limit. Although the perturbation theory for non-Hermitian
Hamiltonian has not been well established, the following result will show
that the corresponding approximation is technically sound by the comparison
with the exact solution. We rewrite the Hamiltonian $H_{\text{eq}}$ in the
form%
\begin{eqnarray}
H_{\text{eq}} &=&H_{\text{0}}+H^{\prime }, \\
H_{\text{0}} &=&\sum_{l=2}^{N-2}\left\vert l\right\rangle \left\langle
l+1\right\vert +\text{H.c.}  \notag \\
&&+\left( V+i\gamma \right) \left\vert 1\right\rangle \left\langle
1\right\vert +\left( V-i\gamma \right) \left\vert N\right\rangle
\left\langle N\right\vert , \\
H^{\prime } &=&\left\vert 1\right\rangle \left\langle 2\right\vert
+\left\vert N-1\right\rangle \left\langle N\right\vert +\text{H.c.,}
\end{eqnarray}%
where the eigen states of $H_{\text{0}}$\ can be easily obtained as $%
\{\left\vert 1\right\rangle $, $\left\vert N\right\rangle $, $%
\sum_{j=2}^{N-1}\sqrt{\frac{2}{N-1}}\sin [\frac{\left( n-1\right) j\pi }{N-1}%
]\left\vert N\right\rangle $; $n\in $ $\left[ 2,N-1\right] \}$ with
corresponding energy $\{V+i\gamma $, $V-i\gamma $, $2\cos [\frac{\left(
n-1\right) \pi }{N-1}]$; $n\in $ $\left[ 2,N-1\right] \}$. This set of eigen
states has a special feature that they can construct a complete set under
the Dirac inner product, even $H_{\text{0}}$\ is a non-Hermitian
Hamiltonian. Then the effective Hamiltonian for two bound states can be
obtained as

\begin{eqnarray}
H_{\text{eff}} &=&\lambda _{\text{eff}}\left\vert 1\right\rangle
\left\langle N\right\vert +\text{H.c.}+(V+V_{\text{eff}}+i\gamma )\left\vert
1\right\rangle \left\langle 1\right\vert  \notag \\
&&+(V+V_{\text{eff}}-i\gamma )\left\vert N\right\rangle \left\langle
N\right\vert ,
\end{eqnarray}%
in the case of $\left\vert V\right\vert \gg 1$, the model above is a simple
two-site model and easily solvable. Here the effective potential is%
\begin{equation}
V_{\text{eff}}=\frac{2}{N-1}\sum_{n=2}^{N-1}\frac{\sin ^{2}\phi _{n}}{%
V-2\cos \phi _{n}}\approx \frac{1}{V},
\end{equation}%
and the effective coupling is%
\begin{eqnarray}
\lambda _{\text{eff}} &=&\frac{2}{N-1}\sum_{n=2}^{N-1}\frac{\sin \phi
_{n}\sin [\left( N-2\right) \phi _{n}]}{V-2\cos \phi _{n}}  \notag \\
&\approx &\frac{\Omega }{V^{2}},
\end{eqnarray}%
where parameters $\Omega $, $\phi _{n}$\ and $\theta $\ are $N$ dependent
functions

\begin{eqnarray}
\Omega &=&\frac{\cos \left[ \left( N-4\right) \pi /2\right] \sin [\left(
N-4\right) \left( N-2\right) \theta ]}{\left( N-1\right) \sin [\left(
N-4\right) \theta ]}  \notag \\
&&-\frac{\left( -1\right) ^{N/2}\sin [\left( N-2\right) N\theta ]}{\left(
N-1\right) \sin (N\theta )}, \\
\phi _{n} &=&2\left( n-1\right) \theta \\
\theta &=&\frac{\pi }{2\left( N-1\right) }.
\end{eqnarray}%
The eigen states of $H_{\text{eff}}$ are%
\begin{equation}
(i\gamma \pm \sqrt{\lambda _{\text{eff}}^{2}-\gamma ^{2}})\left\vert
1\right\rangle +\lambda _{\text{eff}}\left\vert N\right\rangle ,
\end{equation}%
with eigenvalues: $\pm \sqrt{\lambda _{\text{eff}}^{2}-\gamma ^{2}}+V+V_{%
\text{eff}}$. At the EP, $\lambda _{\text{eff}}^{2}=\gamma _{c}^{2}$,\ the
coalescent state is%
\begin{equation}
i\gamma _{c}\left\vert 1\right\rangle +\lambda _{\text{eff}}\left\vert
N\right\rangle ,
\end{equation}%
with energy%
\begin{equation}
\varepsilon _{c}=V+V_{\text{\textrm{eff}}}\approx V+\frac{1}{V},
\end{equation}%
which is in agreement with the approximate expression Eq. (\ref{app c}) in
the Appendix. Then the boundary has the form
\begin{equation}
\ln \left\vert \gamma \right\vert +2\ln \left\vert V\right\vert =\ln
\left\vert \Omega \right\vert ,  \label{App boundary}
\end{equation}%
in the logarithm scales, indicating a linear phase boundary with a fixed
slope. This is qualitatively\ in agreement with the numerical results in
Fig. \ref{fig2}\ obtained by the exact solution, where the slopes of the
boundary are $N$ dependent.

Based on the phase boundary, one can prepare the target state in the
vicinity of the EPs via dynamic process\textbf{.} The target state is a Bell
state, expressed as%
\begin{equation}
\left\vert \text{Bell}\right\rangle =\frac{1}{\sqrt{2}}\left( \left\vert
1\right\rangle -i\left\vert N\right\rangle \right) .
\end{equation}%
The initial state is taken as $\left\vert \psi \left( 0\right) \right\rangle
=\left\vert 1\right\rangle $, the evolved state $\left\vert \psi \left(
t\right) \right\rangle $\ is expected to close the target state for
sufficient long time. We employ the fidelity%
\begin{equation}
f\left( t\right) =\left\vert \left\langle \text{Bell}\right\vert \widetilde{%
\psi }\left( t\right) \rangle \right\vert ,
\end{equation}%
to characterize the efficiency of the scheme. Here $\left\vert \widetilde{%
\psi }\left( t\right) \right\rangle $\ is the Dirac normalized state of $%
\left\vert \psi \left( t\right) \right\rangle $ to reduce the increasing
norm of $\left\vert \psi \left( t\right) \right\rangle $. The time evolution
of the state is computed by numerical diagonalization. For given $N$ and $V$%
, we numerically search an optimal $\gamma $\ to obtain higher fidelity in
the broken symmetric region. In order to quantitatively evaluate the
fidelity and demonstrate the proposed scheme, we simulate the dynamic
processes of the quantum state preparation. To illustrate the process, we
plot the fidelities as functions of time for systems with $N=6$ and $8$ in
Fig. \ref{fig3}. It shows that the fidelities converges to a steady value
exponentially fast. Larger $\left\vert V\right\vert $ corresponds to smaller
optimal $\gamma $, leading to higher fidelity, but longer converging time.
We also find that the converging times for two cases are sensitive to the
size $N$. These accord with the phase diagrams in Fig. \ref{fig2}: linear
boundary indicates that larger ln$\left\vert V\right\vert $ matches smaller
ln$\gamma $\ and slight change of slopes between ln$\left\vert V\right\vert $
and ln$\gamma $ results in drastic change of the converging times.

\begin{figure*}[tbp]
\includegraphics[ bb=23 259 448 579, width=0.45\textwidth, clip]{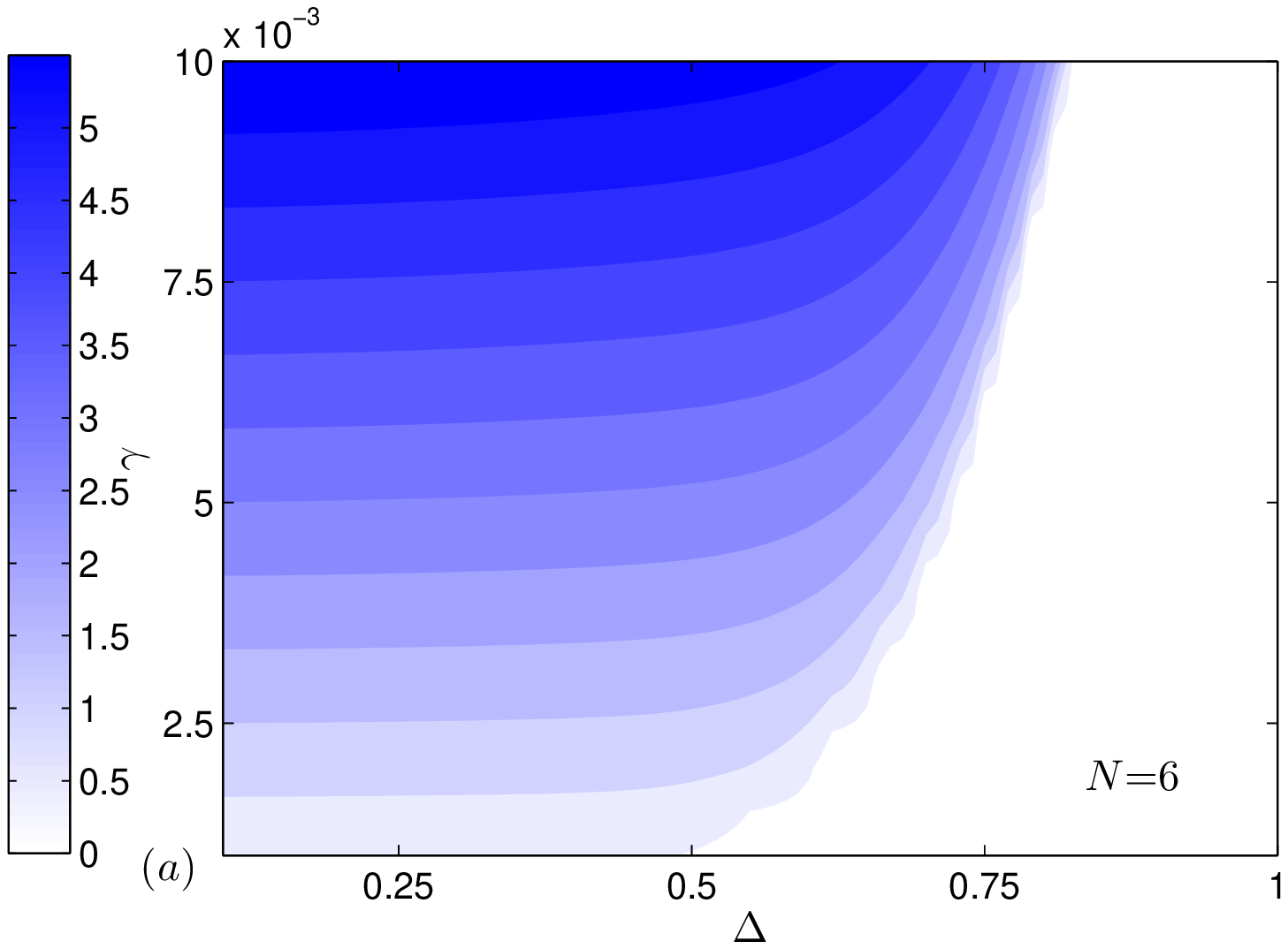} %
\includegraphics[ bb=23 259 448 579, width=0.45\textwidth, clip]{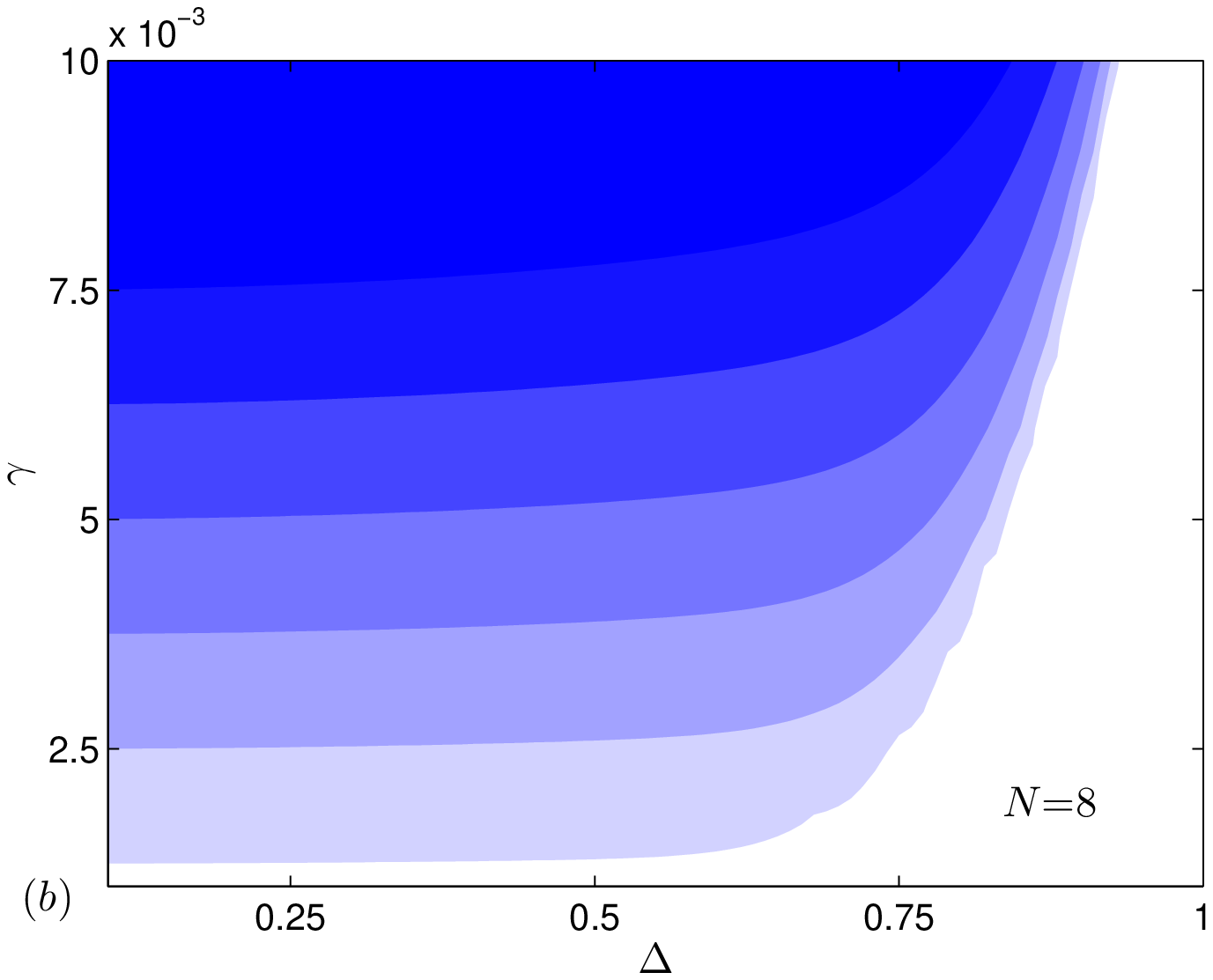}
\caption{(Color online) Phase diagram of non-Hermitian Hamiltonian in Eq. (%
\protect\ref{H_GHZ}). The color contour map represents the magnitude of the
imaginary part of conjugate-pair energy levels for $N=6$ and $8$, obtained
by exact diagonalization. The white area indicates the region where the
spectrum is entirely real. We see that two boundaries have similar shapes
but with a shift.}
\label{fig4}
\end{figure*}

\begin{figure*}[tbp]
\includegraphics[ bb=17 255 436 566, width=0.45\textwidth, clip]{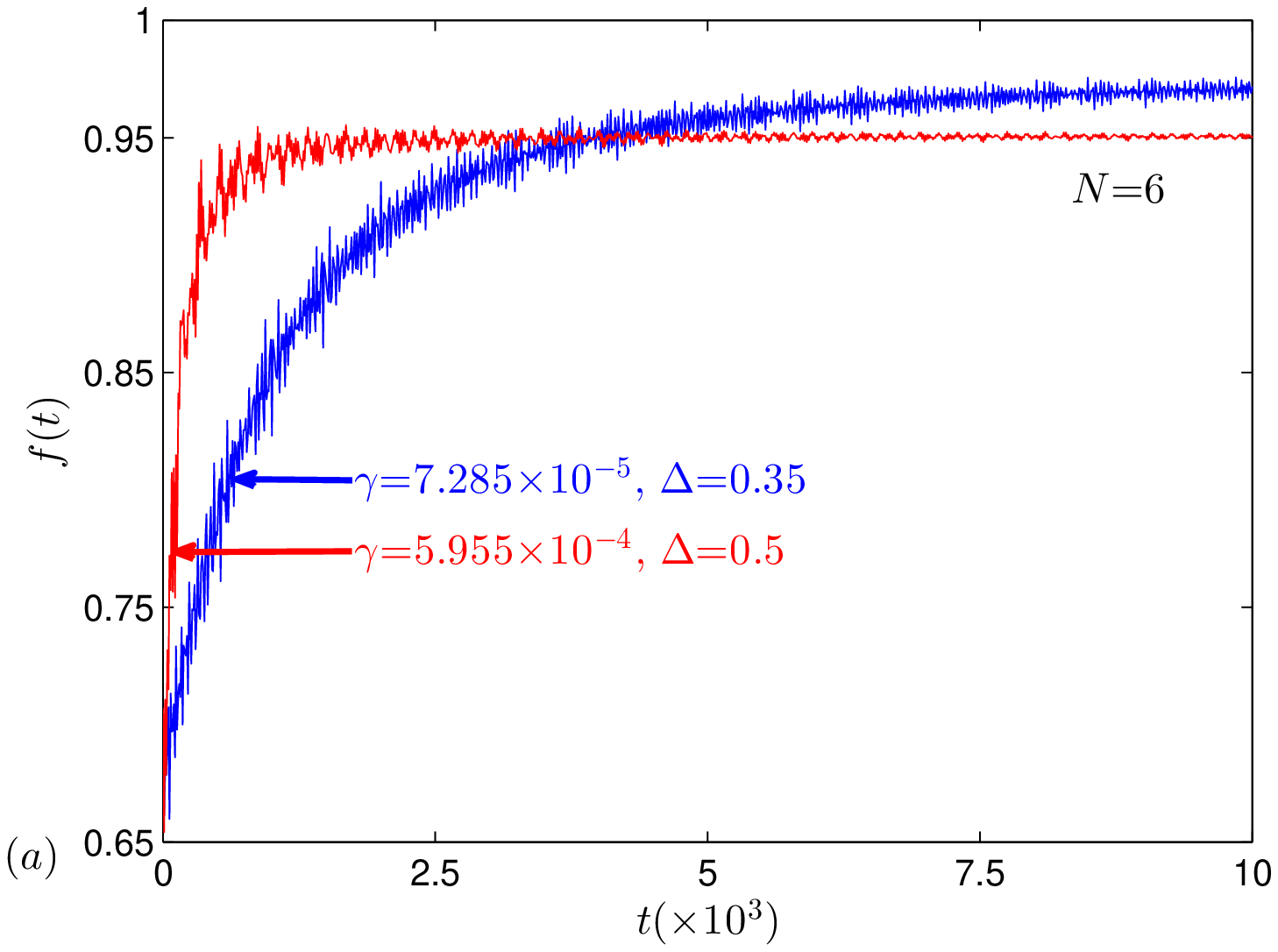} %
\includegraphics[ bb=17 255 436 566, width=0.45\textwidth, clip]{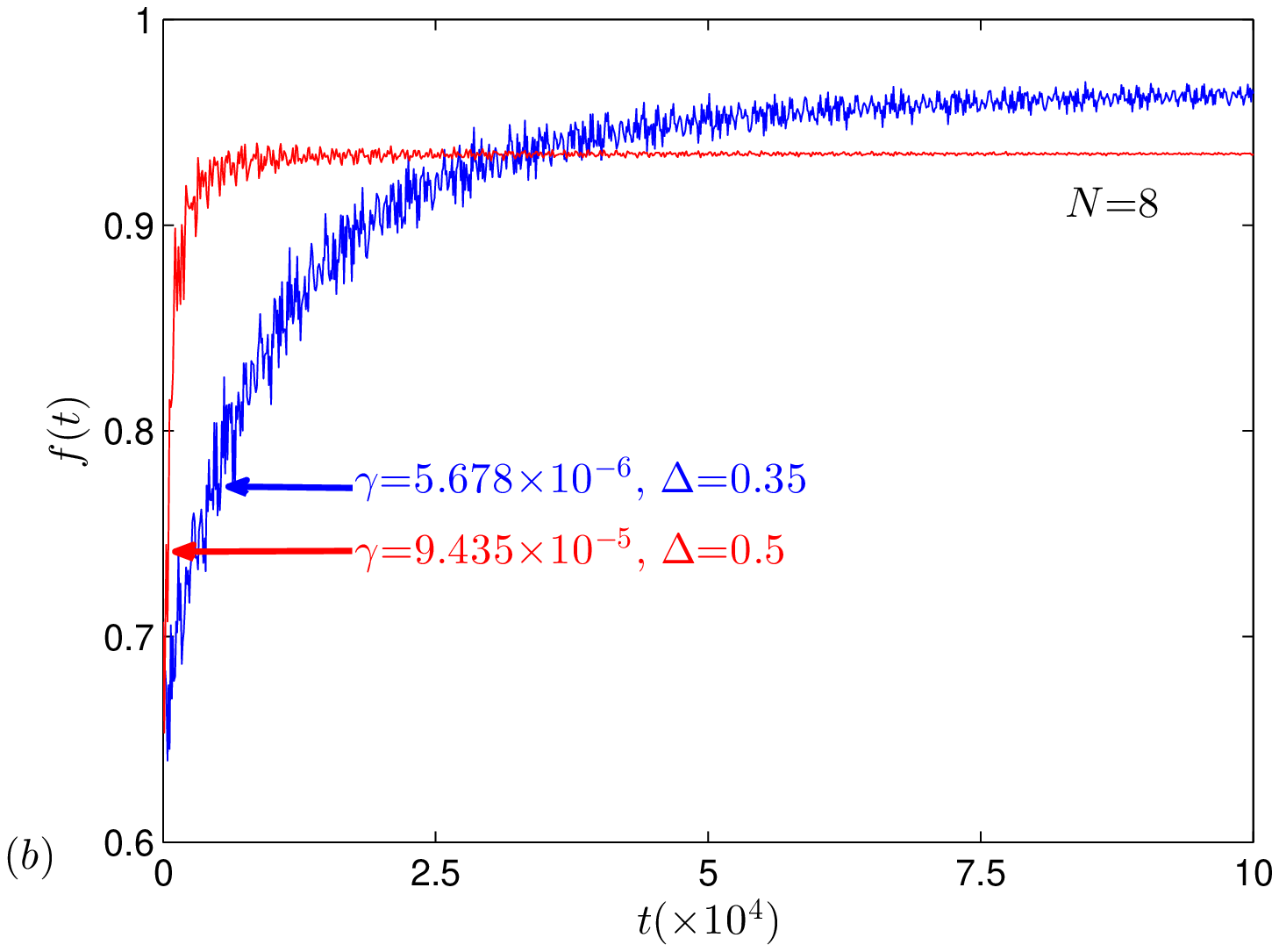}
\caption{(Color online) Plots of the fidelity $f\left( t\right) $ for
preparing GHZ state,\ as a function of time for the systems with $N=6$ (a)
and $8$ (b). The times are dimensionless and in units of $10^{3}$ and $%
10^{4} $,\ respectively.\ We see that the converging fidelities approach to $%
1$, getting higher as decreasing of $\Delta $ and longer time. The time
scale of (b) is over ten times longer than that of (a), which indicates the
difficulty of preparing long-distance GHZ state.}
\label{fig5}
\end{figure*}

\section{GHZ state}

\label{GHZ state}The above conclusion provides a way to prepare a
superposition of two distant position states. Such a scheme can be extended
to prepare the GHZ state which has the form
\begin{equation}
\left\vert \text{GHZ}\right\rangle =\left\vert \Downarrow \right\rangle
+\prod_{l=1}^{N}\sigma _{l}^{+}\left\vert \Downarrow \right\rangle .
\end{equation}%
States $\left\vert \Downarrow \right\rangle $\ and $\prod_{l=1}^{N}\sigma
_{l}^{+}\left\vert \Downarrow \right\rangle $\ can be regarded as two end
position states, which are connected by $N$-step operations of operator $%
\sum_{l=1}^{N}\sigma _{l}^{x}$. This opens a probability to select the GHZ
state as a steady state near the EP. We consider a simple and practical
model, which is a non-Hermitian Ising model, described by the Hamiltonian

\begin{equation}
H_{\text{\textrm{GHZ}}}=-J\sum_{l=1}^{N}\sigma _{l}^{z}\sigma
_{l+1}^{z}+i\gamma \sum_{l=1}^{N}\sigma _{l}^{z}+\Delta \sum_{l=1}^{N}\sigma
_{l}^{x}.  \label{H_GHZ}
\end{equation}%
It is a standard transverse-field Ising model at $\gamma =0$, which can be
exactly solved and has been extensively studied in a variety of areas.
Recently, theoretical studies of several types of quantum Ising models were
extended to the non-Hermitian regime and some peculiar properties were
observed\textbf{\ }\cite{Korff,Castro,Deguchi,Giorgi,Bytsko,Zhang1,Li}%
\textbf{.} In the case of $J=0$, this model is reduced to non-interacting
spin-$1/2$ particles with complex magnetic field, which has full real
spectrum when $\Delta ^{2}\eqslantgtr \gamma ^{2}$ \cite{Zhang3}.\ We assume
that the phase transition can occur in the case of nonzero $\gamma $ and $J$%
. Since this model is not solvable, we perform numerical simulation by exact
diagonalization.

Similarly as last section, we still employ the magnitude of the imaginary
part of the eigen energy $\left\vert \text{Im}\varepsilon \right\vert $ as
an indicator to characterize the phase boundary. Taking $J=1$, we plot $%
\left\vert \text{Im}\varepsilon \right\vert $ as a function of $V$ and $%
\gamma $\ for the systems with $N=6$ and $8$ in Fig. \ref{fig4}. We find
that the phase boundaries of the two cases have the similar profile but with
a shift. This will be reflected on the speed of the fidelity convergence.

For a GHZ state preparation, the initial state is taken as $\left\vert \psi
\left( 0\right) \right\rangle =\left\vert 1\right\rangle $, the evolved
state $\left\vert \psi \left( t\right) \right\rangle $\ is expected to close
the target state for sufficient long time. We employ the fidelity%
\begin{equation}
f\left( t\right) =\left\vert \left\langle \text{GHZ}\right\vert \widetilde{%
\psi }\left( t\right) \rangle \right\vert ,
\end{equation}%
to characterize the efficiency of the scheme. Here $\left\vert \widetilde{%
\psi }\left( t\right) \right\rangle $\ is the Dirac normalized state of $%
\left\vert \psi \left( t\right) \right\rangle $ to reduce the increasing
norm of $\left\vert \psi \left( t\right) \right\rangle $. The time evolution
of the state is computed by numerical diagonalization. For given $N$ and $V$%
, we numerically search an optimal $\gamma $\ to obtain higher fidelity in
the broken symmetric region. In order to quantitatively evaluate the
fidelity and demonstrate the proposed scheme, we simulate the dynamic
processes of the quantum state preparation. To illustrate the process, we
plot the fidelities as functions of time for systems with $N=6$ and $8$ in
Fig. \ref{fig5}. The obtained results are similar to the case of Bell-state
production at last.

\section{Summary}

\label{Summary}In summary, we presented schemes to generate W, distant Bell
and GHZ states by exploiting the quantum phase transitions in non-Hermitian $%
XY$\ and transverse-field Ising spin chains. The phase diagrams for such two
models are obtained analytically and numerically, which is crucial for the
practical realization of the scheme. Numerical simulations on the dynamics
process for state preparation show that the evolved states close to target
states in an exponential manner over time. Comparing the dynamical
preparation of quantum state via Hermitian system, where the acquired state
only emerges within a short time window, this scheme can provide the steady
final state. A shortcoming of the scheme is that the production period for
Bell and GHZ states increases rapidly as cluster size grows. However, this
scheme is more efficient for a W-state production.

\appendix*

\section{Exact solution of the $H_{\text{eq}}$}

In this appendix, we present the exact results for the solutions of
following model

\begin{eqnarray}
H_{\text{\textrm{eq}}} &=&\sum_{l=1}^{N-1}\left\vert l\right\rangle
\left\langle l+1\right\vert +\text{\textrm{H.c.}}  \notag \\
&&+\left( V+i\gamma \right) \left\vert 1\right\rangle \left\langle
1\right\vert +\left( V-i\gamma \right) \left\vert N\right\rangle
\left\langle N\right\vert ,
\end{eqnarray}%
and the EPs in the cases of $V=0$ and $\left\vert V\right\vert >2$.

\subsection{$V=0$ case}

The Bethe Ansatz wave function is in the form%
\begin{equation}
\left\vert k\right\rangle =\sum_{j=1}^{N}\left(
A_{k}e^{ikj}+B_{k}e^{-ikj}\right) \left\vert j\right\rangle ,
\end{equation}%
where $k$ is real number, indicating a scattering state. The Schrodinger
equation $H\left\vert k\right\rangle =\varepsilon _{k}\left\vert
k\right\rangle $ can be written as%
\begin{equation}
M\left[
\begin{array}{c}
A_{k} \\
B_{k}%
\end{array}%
\right] =0,  \label{AB Eqs}
\end{equation}%
where the matrix\
\begin{eqnarray}
M &=&\left[
\begin{array}{cc}
\upsilon _{+}e^{ik}+e^{ik2} & \upsilon _{+}e^{-ik}+e^{-ik2} \\
\upsilon _{-}e^{ikN}+e^{ik\left( N-1\right) } & \upsilon
_{-}e^{-ikN}+e^{-ik\left( N-1\right) }%
\end{array}%
\right] ,  \notag \\
\upsilon _{\pm } &=&\pm i\gamma -\varepsilon _{k},
\end{eqnarray}%
and the real spectrum%
\begin{equation}
\varepsilon _{k}=2\cos k.
\end{equation}%
The existence of solution requires%
\begin{equation}
\det \left\vert M\right\vert =0,  \label{condition}
\end{equation}%
which leads to the equation%
\begin{equation}
F\left( k\right) =\sin \left[ k\left( N+1\right) \right] +\gamma ^{2}\sin %
\left[ k\left( N-1\right) \right] =0.
\end{equation}%
The EP $k_{c}$\ can be determined by equation%
\begin{equation}
F\left( k_{c}\right) =\frac{\partial }{\partial k}F\left( k_{c}\right) =0.
\end{equation}%
We obtain $k_{c}=\pi /2$\ at $\gamma =1$ (see ref. \cite{Jin1}).

\subsection{$\left\vert V\right\vert >2$ case}

In this situation, we are interested in bound states. The corresponding
Bethe Ansatz wave function is in the form%
\begin{equation}
\left\vert \kappa \right\rangle =\sum_{j=1}^{N}\left( \alpha _{\kappa
}e^{\kappa j}+\beta _{\kappa }e^{-\kappa j}\right) \left\vert j\right\rangle
,
\end{equation}%
where $\kappa $\ is a real number. By the similar procedure, we reach the
equation%
\begin{equation}
\digamma \left( \kappa _{c}\right) =\frac{\partial }{\partial \kappa }%
\digamma \left( \kappa _{c}\right) =0,  \label{EP_EQS2}
\end{equation}%
which determines the location of EPs at energy%
\begin{equation}
\epsilon _{\kappa _{c}}=2\cosh \kappa _{c},
\end{equation}%
where function%
\begin{eqnarray}
\digamma \left( \kappa \right) &=&\sinh \left[ \left( N+1\right) \kappa %
\right] -2V\sinh \left( N\kappa \right)  \notag \\
&&+\left( V^{2}+\gamma ^{2}\right) \sinh \left[ \left( N-1\right) \kappa %
\right] .
\end{eqnarray}%
From Eq. (\ref{EP_EQS2}), we have

\begin{eqnarray}
&&\frac{\left[ N\eta _{+}+\eta _{-}\right] \cosh \kappa _{c}-2NV}{\left(
N\eta _{-}+\eta _{+}\right) \sinh \kappa _{c}}  \notag \\
&=&\frac{\eta _{-}\sinh \kappa _{c}}{\eta _{+}\cosh \kappa _{c}-2V}=-\tanh
\left( N\kappa _{c}\right) ,
\end{eqnarray}%
where%
\begin{equation}
\eta _{\pm }=1\pm V^{2}\pm \gamma ^{2}.
\end{equation}%
The bound state EPs require
\begin{equation}
\left\vert \epsilon _{\kappa _{c}}\right\vert >2\left\vert V\right\vert .
\end{equation}%
Such solutions exist when parameters $V$ and $\gamma $ satisfy%
\begin{equation}
(c+\sqrt{c^{2}-1})^{2N}=\frac{\eta _{+}c-2V-\eta _{-}\sqrt{c^{2}-1}}{\eta
_{+}c-2V+\eta _{-}\sqrt{c^{2}-1}},  \label{EPts}
\end{equation}%
which indicates the exact phase boundary and is plotted in Fig. \ref{fig2}
for the cases of $N=6$ and $8$. Here real number $c$ is%
\begin{eqnarray}
&&c=\cosh \kappa =F[1  \notag \\
&&+\sqrt{1-\frac{4N\left( \eta _{+}-1\right) \left( N\eta _{-}^{2}+\eta
_{+}\eta _{-}+4NV^{2}\right) }{V^{2}\left( 2N\eta _{+}+\eta _{-}\right) ^{2}}%
}].
\end{eqnarray}%
\begin{equation}
F=\frac{V\left( 2N\eta _{+}+\eta _{-}\right) }{4N\left( \eta _{+}-1\right) }
\end{equation}%
In the case of $\left\vert V\right\vert \gg 1$, we have%
\begin{equation}
c\approx \frac{V}{2}+\frac{1}{2V},  \label{app c}
\end{equation}%
which gives the approximate energy expression $\epsilon _{\kappa }=2\cosh
\kappa \approx V+\frac{1}{V}$.

\acknowledgments We acknowledge the support of the National Basic Research
Program (973 Program) of China under Grant No. 2012CB921900 and CNSF (Grant
No. 11374163).

\end{document}